\documentclass[conference]{IEEEtran}
\IEEEoverridecommandlockouts

\usepackage{cite}
\usepackage{amsmath,amssymb,amsfonts}
\usepackage{algorithmic}
\usepackage{graphicx}
\usepackage{textcomp}
\usepackage{xcolor}
\usepackage{tikz}
\usepackage{url}

\def\BibTeX{{\rm B\kern-.05em{\sc i\kern-.025em b}\kern-.08em
    T\kern-.1667em\lower.7ex\hbox{E}\kern-.125emX}}
\begin{document}

\title{Sanctorum: A lightweight security monitor for secure enclaves
\thanks{This work was partially funded by Delta Electronics, Analog
Devices, and DARPA \& SPAWAR under contract N66001-15-C-4066, and the DARPA SSITH program under contract
HR001118C0018.}
}

\author{
    \IEEEauthorblockN{Ilia Lebedev$^1$, Kyle Hogan$^1$, Jules Drean$^1$, David Kohlbrenner$^2$, Dayeol Lee$^2$}
    \IEEEauthorblockN{Krste Asanovi\'c$^2$, Dawn Song$^2$, Srinivas Devadas$^1$}
    \IEEEauthorblockA{ $^1$\texttt{\{}ilebedev, klhogan, drean, devadas\texttt{\}}@csail.mit.edu,  \textit{CSAIL, MIT}, Cambridge, MA, USA}
    \IEEEauthorblockA{ $^2$\texttt{\{}dkohlbre, dayeol, krste, dawnsong\texttt{\}}@berkeley.edu,
    \textit{EECS, UC Berkeley}, Berkeley, CA, USA}
}

\maketitle


\begin{abstract}
Enclaves have emerged as a particularly compelling primitive to implement trusted execution environments: strongly isolated sensitive user-mode processes in a largely untrusted software environment.
While the threat models employed by various enclave systems differ, the high-level guarantees they offer are essentially the same: attestation of an enclave's initial state, as well as a guarantee of enclave integrity and privacy in the presence of an adversary.

This work describes Sanctorum, a small trusted code base (TCB), consisting of a generic enclave-capable system, which is sufficient to implement secure enclaves akin to the primitive offered by Intel's SGX.
While enclaves may be implemented via unconditionally trusted hardware and microcode, as it is the case in SGX, we employ a smaller TCB principally consisting of authenticated, privileged software, which may be replaced or patched as needed.
Sanctorum implements a formally verified specification for generic enclaves on an in-order multiprocessor system meeting baseline security requirements, e.g., the MIT Sanctum processor and the Keystone enclave framework.
Sanctorum requires trustworthy hardware including a random number generator, a private cryptographic key pair derived via a secure bootstrapping protocol, and a robust isolation primitive to safeguard sensitive information.
Sanctorum's threat model is informed by the threat model of the isolation primitive, and is suitable for adding enclaves to a variety of processor systems.

\end{abstract}


\vspace{-0.05in}
\section{Introduction}

In order to ensure security for outsourced computations, it is necessary to first consider meaningful security guarantees that are realistically enforceable on a remote platform and how to verify that these guarantees have been upheld. Trusted Execution Environments (TEEs) seek to satisfy these requirements. One of the approaches to TEEs consists of providing strong isolation for user mode processes called \textit{enclaves}. Enclaves are designed to preserve their confidentiality and integrity in the presence of a malicious operating system (OS) or other enclaves. However, defining what it means for an enclave to be truly isolated is not easy when faced with side channel adversaries exploiting leakage from data dependent utilization of shared resources such as caches, OS managed demand paging, or, more recently, speculative execution. 

Hardware support for enforcing low level invariants has been used to provide high level isolation guarantees for systems such as Aegis \cite{suh2003aegis}, SGX \cite{mckeen2013sgx}, Bastion \cite{bastion}, Sanctum \cite{sanctum}, and Komodo \cite{komodo}. Sanctorum, the trusted software enabling the Sanctum processor, focuses on utilizing a small, privileged piece of monitor code in conjunction with partitioned caches and flushing of shared state on context switches to provide isolation for enclaved processes across either space or time. 
This allows enclaves to share hardware resources for performance improvements, but ensures that enclave state cannot be impacted by external code in a data dependent manner either directly or indirectly. This paradigm also allows hardware to evolve separately from the software -- the same hardware can be used with a more powerful security monitor to protect against new attacks, or the security monitor can be used with a different hardware design as long as it supports a minimal set of isolation mechanisms.

\vspace{-0.05in}
\section{Related Work}
Many different approaches for providing trusted execution environments within semi-trusted systems have been proposed \cite{suh2003aegis, bastion, Andronick:2010:TPS:1929004.1929013, komodo}. Differences are primarily found in which components are trusted and whether any measurement or formal verification of these components is provided, whether isolation is provided by hardware or software, and the different adversarial models that are considered.




\paragraph{Hardware Support}
Aegis \cite{suh2003aegis}, Bastion \cite{bastion}, SGX \cite{mckeen2013sgx}, Sanctum \cite{sanctum} and Komodo \cite{komodo} all utilize hardware support to provide isolation for trusted execution environments. Sanctorum is the security monitor for the Sanctum processor.

Komodo and Sanctorum both require isolated memory regions, a protected execution environment for the monitor code, a root of trust for attestations, and a secure source of randomness. Bastion requires more significant changes, including new registers, protected disk regions, and cache modifications, to support its isolation guarantees \cite{bastion}. SeL4 \cite{klein2009sel4} and the work of Andronick et al. \cite{Andronick:2010:TPS:1929004.1929013}, which is based on seL4, do not require specialized hardware and instead rely on the kernel to enforce isolation. 

\paragraph{Formal Verification and Trusted Code Base}
Komodo \cite{komodo}, seL4 \cite{klein2009sel4}, and Andronick et al. \cite{Andronick:2010:TPS:1929004.1929013} all provide formal verification for their isolation models. SGX is neither verified nor easy to inspect as it is primarily implemented as undocumented processor microcode. SeL4 \cite{klein2009sel4} does not provide enclave-like guarantees in terms of isolation, but it does demonstrate an example of large, formally verified software implemented with security as a primary goal.
While Sanctorum itself is not formally verified, its design is based on a formally verified specification for enclaves as described in \cite{subramanyan2017formal}.

\paragraph{Side Channel and Hardware Adversaries}
Komodo \cite{komodo} supports protection against physical attacks on memory while Bastion \cite{komodo} provides defense against physical attacks on memory, busses, and disks. Neither defends against memory access pattern attacks; provable defense against these types of attacks requires Oblivious Random Access Memory (ORAM) as in the Ascend processor \cite{ascend}. Neither Komodo nor Bastion provide any defense against side channel adversaries conducting attacks on shared caches, TLB, etc. SGX and Bastion are also vulnerable to controlled channel attacks in which a malicious OS abuses its control over paging to learn enclave access patterns. Neither seL4 \cite{klein2009sel4} nor  Andronick et al. \cite{Andronick:2010:TPS:1929004.1929013} explicitly consider side channel adversaries, but Andronick et al. mention flushing shared resources before context switches as necessary to enforce isolation.

Sanctum \cite{sanctum} and the current implementation of Sanctorum defend against a large class of side channel attacks but do not consider hardware adversaries. Hardening against an adversary capable of tampering with memory requires the use of the memory controller of the Ascend processor \cite{ascend} in Sanctum.

\vspace{-0.05in}
\section{The Enclave Execution Model}
\label{sec:enclave_execution_model}

An enclave is an isolated process consisting of one or more threads and a exclusive allocation of the machine resources needed by the process: memory (physical pages) and hardware structures (cache lines, etc.). 
Enclaves guarantee integrity and confidentiality for private computations and data in the presence of untrusted privileged software such as an OS or hypervisor~\cite{subramanyan2017formal}.
Therefore, enclaves cannot rely on the OS to transparently provide services without potentially violating the integrity and confidentiality of the enclave's data.

Since an enclave uses isolated resources, neither the direct nor the indirect side effects of an enclave's work are visible to untrusted software, including privileged system software.
However, an enclave is not prevented from deliberately leaking its own secrets, as it is able to access resources shared with it by the OS (in order to receive inputs, interact with I/O, etc), leaking the timing of these accesses.
Although the OS cannot modify any enclave state, it can perform denial of service or wholesale destruction of the enclave, as it orchestrates machine resources.
This, however, does not result in any loss of integrity or confidentiality for the enclave.

In addition to guaranteeing isolated execution and private state, Sanctorum (``the security monitor'', or SM) authenticates enclaves via a cryptographic measurement of their initial state.
These properties allow an enclaved application to implement complex behavior with higher-level security properties.

\vspace{-0.05in}
\section{Threat Model}
\label{sec:threatmodel}

The enclave primitive provides a foundation upon which a secure system can be built.
A prudent software system designer can leverage these isolation containers to construct systems with a small, trustworthy trusted computing base.

SM assumes an insidious privileged software adversary able to subvert any software (other than SM) in order to impersonate, tamper with, or inspect an enclave. Denial of service is impossible to defend against in this setting, and is therefore not considered (SM \emph{does} ensure an OS is able to stop a runaway enclave). The specific abilities of the modelled adversary to inspect enclaves via indirect means, such as cache tag state or availability of other shared resources, depends on the availability of isolated protection domains for these surfaces in the hardware platform.

\vspace{-0.075in}
\subsection{Trust assumed by SM}
\label{sec:threatmodel:trust}

A given enclave binary is assumed trustworthy, but is authenticated via a measurement of its initial state taken by SM.
SM's binary image is also assumed to be trustworthy (but is authenticated via a secure boot protocol and endowed with unique keys~\cite{secure_boot}), and is covered by the attestation.
The hardware platform is unconditionally trusted, as hardware defies cryptographic measurement, and should be authenticated as part of remote attestation.
Trust in the authenticated binaries can be garnered through formal verification, rigorous testing, etc.
A trusted first party is required to verify remote attestations (cf. Section \ref{sec:attestation:remote}) on enclave state.
This process requires a PKI to bootstrap trust in the hardware and SM.

Enclaves are trusted to neither compromise their own integrity nor transmit private state to a potential adversary, e.g., by copying it to shared memory.
An enclave's interactions with other software, including SM, may transmit information; the enclave is trusted not to perform these communications in ways dependent on private information.
Any such communication leaks at a minimum the timing of this communication, and may further leak information about microarchitectural state influenced by an enclave's private execution.
The text of the enclave binary is trusted to use communication judiciously, and block or tolerate any leakage permitted by the hardware platform that is within the threat model.
The hardware platform is trusted to respect the text of the enclave and not spontaneously (e.g., speculatively) perform operations that transmit information across protection domains.

SM maps the high-level semantics of enclaves to low-level machine configuration to enforce isolation of machine resources along protection domain boundaries.
To accomplish this, SM checks and maintains that the untrusted system software's allocation of machine resources to enclaves respects protection domain boundaries and is exclusive.
SM relies on the underlying hardware platform to implement meaningful isolation of the resources across protection domains.
Side channel leakage, in particular, requires strong isolation by the hardware platform either by flushing state between context switches or by ensuring that the resource is not shared by different protection domains. SM requires the underlying hardware to guarantee several properties outlined below.

\vspace{-0.05in}
\subsection{Hardware Platform Requirements}
\label{sec:threatmodel:assumptions}

In order to achieve secure enclaves, Sanctorum requires several high-level properties of the underlying hardware platform:

\subsubsection{Memory isolation across protection domains}
\label{sec:threatmodel:assumptions:isolation}

The hardware must guarantee isolation of ``protection domains'' in order to allow SM to isolate itself from any other piece of software and enclaves from each other and from the untrusted OS in memory.
The hardware platform must also be able to restrict access by external actors: SM must be able to restrict DMA by devices to memory owned by SM or enclaves.

\subsubsection{Isolated computation across protection domains}

The hardware platform must guarantee isolation for all the shared resources considered by the threat model. These resources are partitioned across protection domains (if the hardware can support this) with non-interference across partitions, or time-multiplexed across protection domains and cleaned by SM at each re-allocation. For example, the MIT Sanctum processor time-multiplexes cores (including register files and all microarchitectural state and L1 caches), partitions the shared L2 cache (via page coloring), and excludes the shared coherence and DRAM bandwidth from consideration in the threat model. Other platforms may choose to protect other surfaces, which will affect the platform's threat model.

\subsubsection{Exclusive elevated privilege for SM}
\label{sec:threatmodel:assumptions:pki}

In order to prevent efforts by the OS or other software to supplant SM execution and violate security invariants, SM must execute at a higher privilege level than any untrusted software, and have exclusive unrestricted access to physical memory. The hardware must support such privilege. The SM must also be able to interpose on hardware events such as faults and interrupts, as these events may cause a change in the protection domain on whose behalf a core executes, and require machine resources be cleaned and re-allocated. For example, the OS must not be able to execute its fault handler on a core with enclave permissions by sending a software interrupt; SM must be able to receive the interrupt, perform an enclave exit on the core, and then delegate the interrupt to the OS.

\subsubsection{Cryptography  for attestation}
Enclaves must have private access to a trusted source of entropy to perform key agreement and seed cryptographic keys. The hardware platform must enable a trusted public key infrastructure (PKI) for SM (a secret attestation key backed by certificates conveying trust in this SM on this hardware platform) which enables remote parties to authenticate the hardware and the measurement root.

\vspace{-0.05in}
\section{Security Monitor}
\label{sec:monitor}

Sanctorum (SM) implements a small, trusted, privileged security monitor in order to enforce a security policy over the untrusted system software's handling of machine resources.
SM translates the high-level semantics of isolated enclaves into the low-level isolation invariants of machine resources along protection domain boundaries, as implemented by trusted hardware~(cf. Section \ref{sec:threatmodel:assumptions:isolation}).
SM is not a kernel, as it does not make resource management decisions, instead only \emph{verifying} the decisions made by system software.

\vspace{-0.05in}
\subsection{Security Monitor Interface}
\label{sec:monitor:api}

\begin{figure}[ht]
  \vspace{-10pt}
  \caption{SM API via system exceptions, much like a system call.}
  \label{fig:api}
  \begin{center}
    \includegraphics[width=0.48\textwidth]{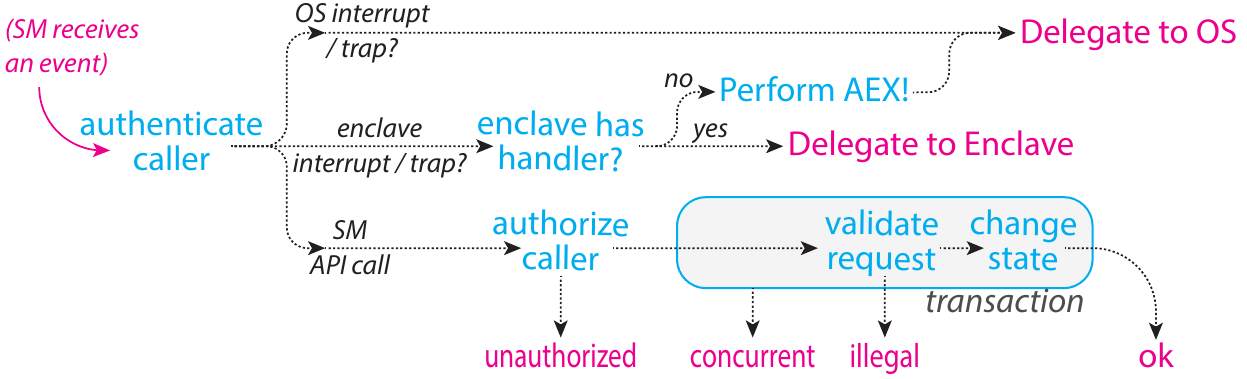}
  \end{center}
  \vspace{-16pt}
\end{figure}

SM implements an API for enclaves and untrusted system software to indirectly manage system resources, as permitted by SM's security state machine.
SM also interposes on machine events such as page faults and interrupts in order to ensure that these events do not violate the system's security policy.
SM's API calls are also implemented via machine events as a system call to SM.
As shown in Figure~\ref{fig:api}, the interface forwards OS events to the OS handler, but requires an Asynchronous Enclave Exit, or AEX~(cf. Section \ref{sec:monitor:enclaves}) to clean sensitive processor state before delegating the event to the OS.
Enclaves can implement fault handlers, and receive some traps/faults in order to implement paging or handle some exceptions.
The OS is always able to de-schedule an enclave by interrupting it, forcing an AEX.

The SM API is highly concurrent on a multicore processor, and requires transaction semantics for most API calls.
After authorizing the caller, SM uses fine-grained locks, and fails transactions in case of a concurrent operation.
SM checks the API call against the machine's current security policy to ensure SM cannot be asked to violate an enclave, nor allow a malicious enclave to compromise the untrusted system.

\vspace{-0.075in}
\subsection{Machine Resources managed by SM}
\label{sec:monitor:resources}

SM enforces invariants over the system software's allocation of isolated resources (cores, physical memory, cache lines, etc.) to their respective protection domains (SM itself, untrusted software, and each enclave).
Protection domains must be non-overlapping with respect to machine resources in order to guarantee isolation: sharing of resources leaks their availability, and allows indirect observation of private information (as generously exemplified by the recent proliferation of cache timing attacks).
Furthermore, protection domains must not be allowed to modify resource allocations of other protection domains: a malicious OS would remove a portion of enclave's memory, and learn private information if the enclave generates a fault.

\begin{figure}[ht]
  \vspace{-10pt}
  \caption{Generic resource state transitions enforced by SM.}
  \label{fig:resources}
  \begin{center}
    \includegraphics[width=0.3\textwidth]{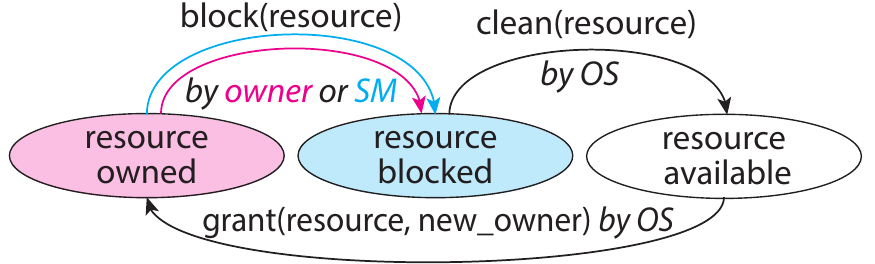}
  \end{center}
  \vspace{-10pt}
\end{figure}

This does not mean enclaves are static.
Instead, an enclave may collaborate with the OS to implement dynamic behaviors like re-allocation of resources or time-multiplexing of existing resources (e.g., demand paging).
As shown in Figure~\ref{fig:resources}, a protection domain can block (\verb|block_resource(type, rid)|) a resource it owns, which the OS will be able to reclaim or re-allocate to a new owner by cleaning it (\verb|clean_resource(type, rid)|).
An existing domain can accept (\verb|accept_resource(type, rid)|) resources the OS offers, completing the transition of the resource to a new protection domain.

SM maintains a map of each resource to its respective owner and a lock via \emph{resource metadata}.
Metadata arrays for statically partitioned resources (e.g., cores, static memory and cache partitions) are pre-allocated as part of SM's binary image.
The management of dynamic resources (e.g., enclaves, threads, and intervals of physical memory, if applicable) is implementation-specific (cf. Section~\ref{sec:implementation}); the metadata must wholly reside in SM-owned memory, and be non-overlapping with other structures.
The management of all mutable resources takes place indirectly via calls to SM's narrow API by the resource owner or the untrusted system software, within SM's security invariants.
SM also maintains some global static state, such as the expected measurement (cf. Section~\ref{sec:attestation:measurement}) of the signing enclave (cf. Section~\ref{sec:attestation:remote}), and SM's certificates and keys.

\vspace{-0.075in}
\subsection{Enclaves and Enclave Threads managed by SM}
\label{sec:monitor:enclaves}

SM implements enclaves: strongly isolated processes with guarantees of exclusive access to a set of machine resources.
At a minimum enclaves use private physical memory containing enclave private virtual pages and page tables, with additional isolation (cache lines, etc.) if implemented by the hardware platform.

Enclave metadata tracks various properties (the enclave's measurement, virtual range, lifecycle state, lock), thread IDs (\verb|tid|), and the machine resources owned by this enclave.
The metadata also contains mailboxes (cf. Section~\ref{sec:attestation:local}) used for trusted inter-enclave communication.
While SM authenticates enclaves via their measurement (cf. Section~\ref{sec:attestation:measurement}), enclave IDs (\verb|eid|) are used to refer to the enclave data structure throughout the SM API.
An \verb|eid| is the physical address of the enclave's metadata structure. SM and untrusted software are identified via reserved constants.

Enclaves use private page tables for accesses within the enclave virtual range (\verb|evrange|), and manage their own private memory, as needed.
Accesses to memory shared with the operating system (outside \verb|evrange|) are implementation-dependent, and may leak timing information.

\begin{figure}[ht]
  \vspace{-10pt}
  \caption{Enclave state transitions enforced by SM.}
  \label{fig:enclave}
  \begin{center}
    \includegraphics[width=0.45\textwidth]{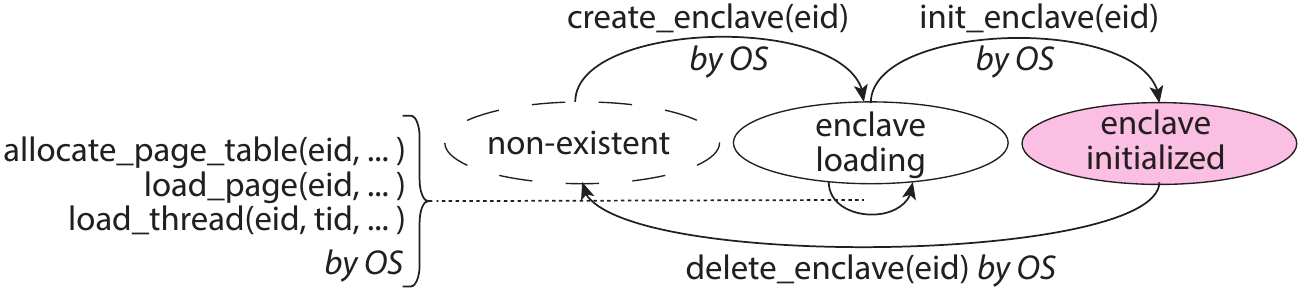}
  \end{center}
  \vspace{-12pt}
\end{figure}

The lifecycle of an enclave is illustrated in Figure~\ref{fig:enclave}: untrusted system software creates an enclave, via \verb|create_enclave(eid, evrange, resources...)|, in a free segment of SM-owned memory (SM enforces safety). After the enclave metadata is created, the OS can grant memory and other resources to the newly created \verb|eid|.
Further, SM can initialize the enclave's physical and virtual memory by reserving space for enclave-private page tables, copy pages from untrusted memory to the enclave's virtual memory, and create threads (\verb|allocate_page_table(...)|, \verb|load_page(...)|, \verb|create_thread(...)|, respectively (not detailed here for brevity).
A \verb|init_enclave(eid)| API call ``seals'' the enclave, preventing further modifications by untrusted software via the API, finalizing the measurement, and allowing the enclave's threads to be scheduled on a core (as described below).
An enclave can block its own resources, or accept new resources granted by the OS, leaking the timing of these operations.
The untrusted system software can destroy an enclave in its entirety, blocking all of its owned resources (\verb|delete_enclave(eid)|), provided none of its threads are scheduled.
SM will require all of the enclave's resources be cleaned~(cf. Section~\ref{sec:monitor:resources}) before they can be re-allocated.

Enclave threads scheduled onto processor cores (via \verb|enter_enclave(eid, tid)|) will execute uninterrupted until either the enclave exits (\verb|exit_enclave()|), or an event causes an asynchronous enclave exit (AEX), e.g., as a result of the OS de-scheduling the enclave.
Upon an AEX, SM saves the state of the enclave thread being suspended into a reserved AEX state structure in the thread metadata, and sets a flag indicating that an AEX had occurred. If the enclave re-enters, it will execute from its entry point, but may respond to the presence of the AEX state to resume execution, if implemented by the enclave.
Before delegating execution to the OS, SM cleans the core's state (this is a re-allocation of the ``core'' resource to another protection domain).

\begin{figure}[ht]
  \vspace{-10pt}
  \caption{Enclave thread state transitions enforced by SM.}
  \label{fig:thread}
  \begin{center}
    \includegraphics[width=0.48\textwidth]{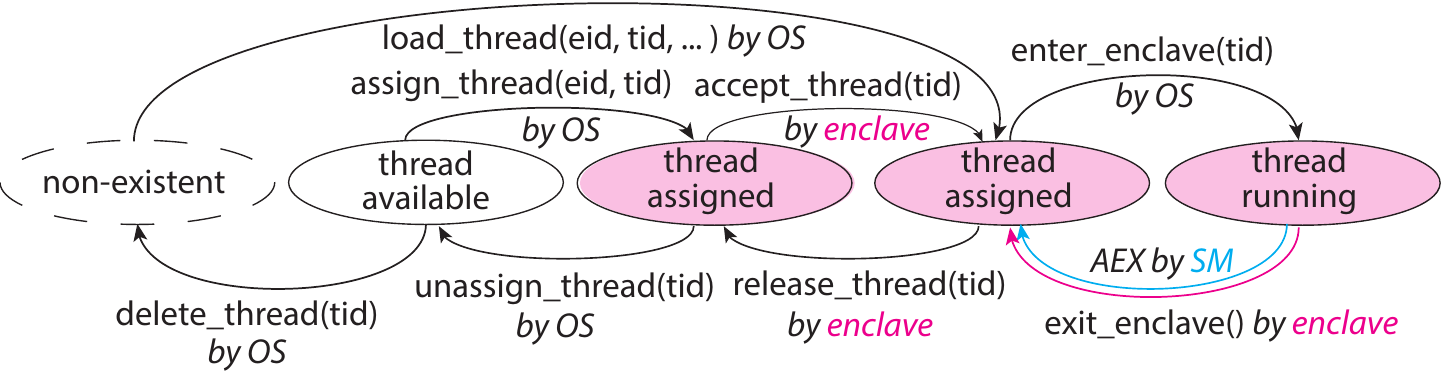}
  \end{center}
  \vspace{-12pt}
\end{figure}

Thread metadata structures are another first-class type recognized by SM, their lifecycle illustrated in Figure~\ref{fig:thread}.
Like enclave metadata, the physical address of a thread's metadata is a thread ID (\verb|tid|), and is used to refer to the thread throughout the API.
The thread metadata tracks the thread's owner enclave, lock, the core it is scheduled on, the presence of an AEX state dump, and the address to execute upon \verb|enclave_enter|, as well as the addresses of fault handlers. Thread metadata also reserves space for core state in case of an AEX and, separately, in case of a fault. After a thread is created, it is assigned to an enclave. Once the enclave is destroyed or blocks the thread, it can be cleaned and re-allocated to another enclave. The enclave can accept the thread via \verb|accept_thread(tid)|.

\vspace{-0.05in}
\section{Enclave Attestation}
\label{sec:attestation}

Attestation allows enclaves to prove their authenticity to local or remote parties leveraging trust in SM, the processor, and the manufacturer's PKI. SM provides a trusted message-passing interface for local attestation of enclaves, and trusts a specific ``signing'' enclave to produce certificates for remote attestation with SM's secret key. A trusted first party is expected to verify this certificate to ascertain trust in the initial state of the enclave being attested to.

\vspace{-0.075in}
\subsection{Measurement}
\label{sec:attestation:measurement}

SM measures enclaves via a sha3~\cite{sha3} cryptographic hash computed for each enclave as part of initialization.
This measurement covers the enclave's configuration, private virtual memory, and any global state necessary to convey trust (e.g., the identity of SM and capabilities of the hardware).
SM performs all operations affecting the initial state of the enclave, and thus has sufficient authority to compute the measurement.
Each operation performed by SM on behalf of the OS as part of enclave initialization (creating the enclave data structure, reserving space for page tables, loading pages, loading threads) extends the enclave's hash with each operation to produce a final measurement at initialization.

Two equivalent enclaves initialized with identical virtual addresses will have equal measurements; the physical addresses used when initializing the enclave are not covered by measurement.
In order to ensure that measurement is descriptive of the enclave's initial state, the mapping between an enclave's virtual page numbers and pages in physical memory must be injective (no aliasing).
To simplify SM's logic needed to enforce this invariant, SM requires that enclaves be loaded in ascending (monotonically increasing) order of physical page numbers.
The enclave's page tables must be initialized before any data, and are at the base of its physical address space.

The measurement of an enclave's initial state authenticates the enclave, provided the enclave, SM, and hardware platform maintain the enclave's integrity after measurement.
This authentication is a necessary part of attestation, which conveys trust in a local or remote enclave to a party (conditional on trust in the hardware, SM, and enclave measurement).

\vspace{-0.075in}
\subsection{Local Attestation}
\label{sec:attestation:local}

In the case where both the enclave being attested to and the verifying enclave execute on the same hardware platform, under the same SM, local attestation is available.
SM guarantees integrity and sender identity for local messages without cryptographic proofs through its authority over all other software: by implementing a trusted, authenticated message passing API, local enclaves can prove their identity to other local enclaves via their mutual trust in SM.

\begin{figure}[ht]
  \vspace{-10pt}
  \caption{Mailbox state transitions enforced by SM.}
  \label{fig:mailbox}
  \begin{center}
    \includegraphics[width=0.35\textwidth]{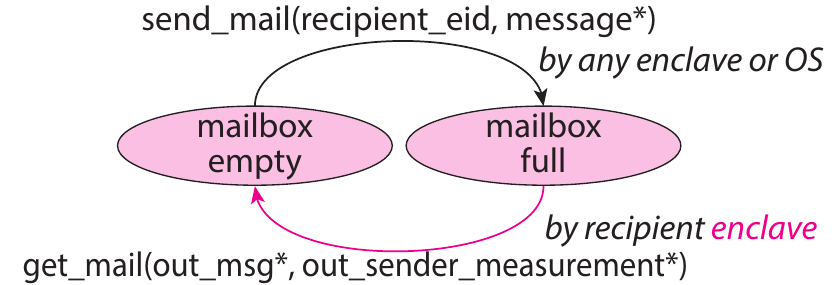}
  \end{center}
  \vspace{-16pt}
\end{figure}

Specifically, SM endows each enclave metadata structure in SM memory with a buffer of one or more ``mailboxes'' used by that enclave to receive authenticated messages. As shown in Figure~\ref{fig:mailbox}, each mailbox can receive mail tagged with the measurement of the sender by SM via SM's \verb|send_mail(recipient_id, message)|, \verb|get_mail(sender_id, out_msg*, out_sender*)| APIs. In order to thwart denial of service by a malicious sender, the recipient must signal their intent to receive from a specific sender via the \verb|accept_mail(sender_id)| API.

\begin{figure}[ht]
  \vspace{-10pt}
  \caption{A local attestation of $E_1$ by $E_2$.}
  \label{fig:attestation_local}
  \begin{center}
    \includegraphics[width=0.4\textwidth]{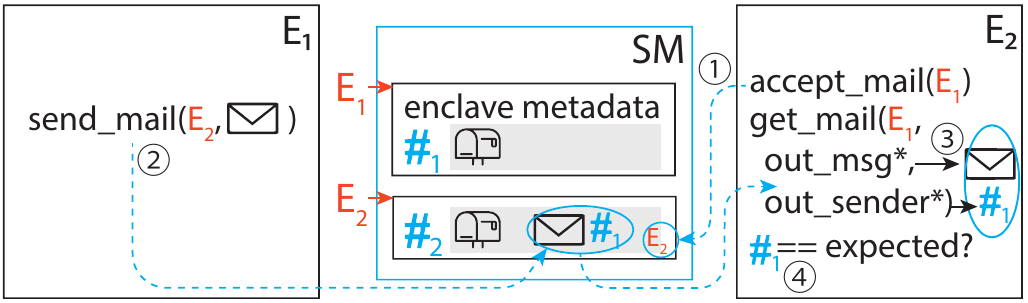}
  \end{center}
  \vspace{-10pt}
\end{figure}

Consider the example in Figure~\ref{fig:attestation_local}.
Here, Enclave $E_2$ attests enclave $E_1$, and untrusted system software informs the participating enclaves of the relevant sender IDs.
$E_2$ signals its intent to receive messages from $E_1$~\textcircled{1}, which enables E1 to send a message to $E_2$~\textcircled{2}. SM stores the message in $E_2$'s mailbox for communication with $E_1$.
SM also records the sender's measurement. The recipient, $E_2$, fetches its messages~\textcircled{3}, and can validate the sender's hash against an expected constant~\textcircled{4} in order to authenticate the message.

\vspace{-0.075in}
\subsection{Remote Attestation}
\label{sec:attestation:remote}

Attestation without a trusted communication medium requires cryptography.
In order to provide enclave attestations to a remote verifier, SM relies on a remote attestation protocol to establish a private channel, present a certificate connecting the enclave to a root of trust, and sign a nonce provided by the verifier.
Specifically, a key agreement scheme derives a shared key for encrypted communication without trust in the system software or network. (If attestation succeeds, the remote party relies on enclave isolation to safeguard the shared key.)
Recall that SM assumes (cf. Section~\ref{sec:threatmodel:assumptions:pki}) a trusted signing key and PKI able to connect trust in SM and its hardware platform.
SM produces an attestation via this signing key by signing an enclave's message and measurement, but does not itself guarantee a confidential execution environment (because SM itself is a shared resource), relying instead on a trusted ``signing enclave'' to compute the signature.
The signing enclave's measurement is hard-coded in the security monitor, allowing it to retrieve the key via a \verb|accept_mail(get_key)| API, while local attestation~(cf. Section~\ref{sec:attestation:local}) allows the enclaves seeking attestation to communicate with the signing enclave.
SM also stores the certificate(s) needed to ascertain its trustworthiness via the trusted PKI, and exposes them via a public API \verb|get_field(field_id, out_data*)|.

\begin{figure}[ht]
  \vspace{-10pt}
  \caption{A remote attestation of E1 by a trusted first party}
  \vspace{-6pt}
  \label{fig:attestation_remote}
  \begin{center}
    \includegraphics[width=0.49\textwidth]{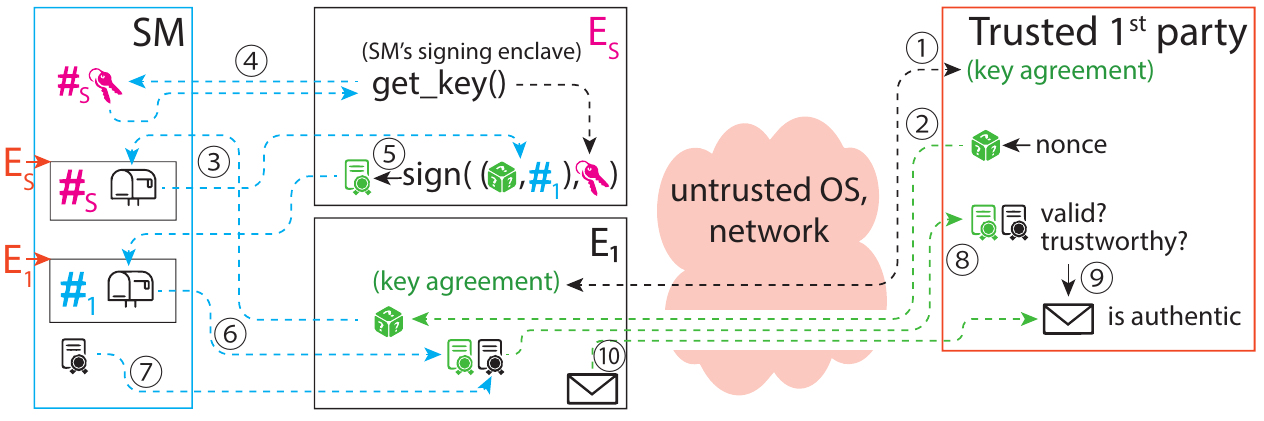}
  \end{center}
  \vspace{-15pt}
\end{figure}

Consider an attestation of enclave $E_1$ by a trusted remote first party exemplified in Figure~\ref{fig:attestation_local}.
The OS is responsible for scheduling the signing enclave $E_S$, communicating relevant enclave IDs, and providing (untrusted) I/O to the trusted first party.
After key agreement~\textcircled{1}, the remote party generates a nonce~\textcircled{2}, which $E_1$ sends to $E_S$ via a mailbox ~\textcircled{3}. $E_S$ fetches SM's key~\textcircled{4} and signs the nonce and $E_1$'s measurement to produce an attestation~\textcircled{5}. $E_1$ receives its attestation via its mailbox~\textcircled{6}, and assembles a message to the remote party: SM's certificate~\textcircled{7} cryptographically connects the attestation to the trusted PKI.
The remote party must receive~\textcircled{8} and verify~\textcircled{9} the attestation in order to bootstrap trust in the encrypted channel created via key agreement.
Provided the attestation succeeds, the shared key authenticates all subsequent messages~\textcircled{10} sent by $E_1$.

\vspace{-0.05in}
\section{Architecture-specific components}
\label{sec:implementation}

SM (cf. Section~\ref{sec:monitor}) implements a monitor able to support enclaves on an abstract machine consisting of an array of typed resources isolated by the hardware platform, including, at a minimum, cores and physical memory.
Refining the high-level tasks of cleaning resources and assigning them to protection domains is specific to the hardware platform.
Of importance is SM's implementation of memory: private segments of physical memory are used throughout SM, but SM does not prescribe specific means by which memory is isolated.

\vspace{-0.075in}
\subsection{MIT Sanctum processor}
\label{sec:implementation:sanctum}

In the MIT Sanctum Processor~\cite{sanctum}, memory isolation is provided by allocating memory in the form of 64 isolated DRAM regions of fixed size (32 MB).
SM for Sanctum straightforwardly stores dynamic arrays in ``metadata regions'': SM-owned regions granted to it by the OS.
DRAM regions are isolated throughout the shared memory hierarchy including the last-level cache.
A page table walk invariant guarantees TLB entries conform to the allocation DRAM regions, requiring a TLB shootdown whenever DRAM regions are re-allocated to a different protection domain.

A small set of hardware modifications over a baseline RISC-V processor implement physical isolation of SM memory from all software, a private page walk for addresses within \verb|evrange|, and enforce physical memory permissions at the granularity of DRAM regions.
RISC-V's M-mode straightforwardly grants SM ultimate authority and access in a Sanctum processor system.
A secure boot protocol~\cite{secure_boot} endows SM with keys rooted in its measurement and the specific device.
Sanctum's cores are in-order, single-thread pipelines, and are exclusively scheduled to protection domains.
When cleaned, the processor flushes its private caches and architected state.

SM is largely implemented in portable, modular C99 code for simplified verification.
The existing implementation for the MIT Sanctum processor consists of 5785 LOC (C: 5264 LOC, Assembly: 521 LOC).
Much of this code is a cryptographic hash function, standard C library functions, and privileged code required to boot a modern OS.
Excluding these, the non platform-specific SM code weighs in at 1011 LOC of C99.

\vspace{-0.05in}
\subsection{Keystone Enclave Framework}
\label{sec:implementation:keystone}

Keystone \cite{keystone} is an enclave framework using RISC-V's powerful physical memory protection (PMP) primitive \cite{riscv-priv-1-10}, and does not rely on hardware modifications to standard RISC-V processors. PMP allows dynamic white-listing of intervals of memory as being accessible by specific privilege modes. Keystone contains an independent implementation of Sanctorum concepts to meet the same objectives using PMP. 

For memory isolation, SM straightforwardly marks its own private state as solely accessible via RISC-V's M-Mode, allowing the OS to access physical memory outside of this forbidden range, and granting itself unrestricted access. Enclaves are likewise marked via a white-listed range of physical memory of arbitrary size. Enclaves use a private set of page tables for all memory accesses, and both these tables and the memory region are protected by PMP. For access to shared resources outside \verb|evrange| the enclave has a shared memory section in its page tables mapped to an OS-allocated untrusted buffer. Keystone does not, at the time of this writing, isolate microarchitectural resources such as shared cache lines across arbitrary platforms, as reflected by its threat model.

\vspace{-0.05in}
\section{Conclusion}
\label{sec:conclusion}

Sanctorum (SM) is a minimal security monitor for enclaved computations running on in-order multiprocessors. It enforces a set of low level isolation properties to provide confidentiality and integrity for remote computations. Sanctorum prevents realistic side channel attacks against shared caches and attacks on demand paging. Sanctorum is being expanded to include proposed defenses against recently discovered attacks on speculative execution such as Spectre.

\vspace{-0.05in}

\bibliographystyle{abbrv}
\bibliography{biblio}

\end{document}